\documentclass[aps,preprint,showpacs,amsmath,amssymb]{revtex4-1}
\usepackage{graphicx}
\usepackage{dcolumn}
\usepackage{bm}
\usepackage{hyperref}
\usepackage{xcolor}
\usepackage{soul}
\begin{document}

\title{Geometric Characterization and Feasible Region Analysis of Bipartite Qutrit Bound Entanglement}
\author{Mazhar Ali \footnote{Email: mazharaliawan@yahoo.com or mazhar.ali@iu.edu.sa}}
\affiliation{Department of Electrical Engineering, Faculty of Engineering, Islamic University of Madinah, 
107 Madinah, Saudi Arabia}


\begin{abstract}
We characterize bound entanglement in bipartite qutrit systems by combining the Positive Partial Transpose (PPT) criterion with a structural, non-completely positive map. Focusing on a comprehensive three-parameter bipartite qutrit family $\rho_{a,b,c}$, we present an exact analytical and geometric partitioning of the physical state space simplex. We derive the closed-form quadratic boundary governing the leaf-like PPT region $(a^2+ab+b^2-b \leq 0)$ over the domain $0 < a \leq \frac{1}{3}$, and identify the exact linear threshold $(a > c)$ that characterizes the PPT bound entangled sub-region. This family strictly generalizes the well-known Horodecki bound entangled states, which appear as a single slice of the three-dimensional leaf. Applying the structural map to this decomposition, we show that it certifies bound entanglement throughout the entire threshold region, accounting for $14.76\%$ of the total PPT leaf area.
\end{abstract}

\pacs{03.65.Db, 03.65.Ud, 03.67.-a}

\maketitle

\section{Introduction}\label{S-intro}

Quantum entanglement not only defies our classical intuition about correlations between physical systems, but it also underlies several emerging quantum technologies \cite{Erhard-NRP-2020,Friis-2019}. Consequently, the characterization and detection of entanglement has remained an active and central problem in quantum information theory over the last few decades \cite{Horodecki-RMP81,Guehne-PR474,Eisert-RMP82}. Despite this progress, determining whether a general quantum state is entangled or separable is still a hard problem. For bipartite systems with $\dim \, \mathcal{H}_{AB} \leq 6$, the {\it Peres-Horodecki criterion} provides a necessary and sufficient test: a state is separable only if the partial transpose of its density matrix remains positive (PPT) \cite{Peres-PRL77,Horodecki-PLA223}. In higher dimensions, however, PPT is only a necessary condition, and there exist entangled states that nevertheless remain positive under partial transposition. Such states are termed {\it bound entangled}: no pure entanglement can be distilled from them by local operations and classical communication, even though they are provably non-separable \cite{Horodecki-PRL82}. The realignment criterion \cite{Rudolph-QIP4,Chen-QIC3} can detect some of these PPT entangled states, but it fails for many others. The problem is even richer for multipartite systems, where fully separable, biseparable, fully inseparable, and genuinely entangled states must all be distinguished \cite{Schneeloch-PR2020,Jungnitsch-PRL106,Novo-PRA88,Hofmann-JPA47,Guehne-NJP12,Guehne-PLA375,Bergmann-JPA46,Zhou-NPJ5,Xu-PRA107}.

Because full quantum state tomography becomes increasingly demanding as system size grows, several measurement-efficient alternatives have been proposed to certify entanglement without reconstructing $\rho$ completely. These include criteria based on moments of the partially transposed density matrix (PT-moments) \cite{Paris-LNP-2004,Tran-PRA92-2015,Tran-PRA94-2016,vanEnk-PRL108-2012,Elben-PRA99,Brydges-Sc364,Elben-PRL125,Huang-NP16,Yu-PRL127,Neven-npjQI7}, protocols based on local randomized measurements \cite{Ketterer-PRL122,Ketterer-Q4,Ketterer-PRA106,Ohnemus-PRA107,Wyderka-PRL131,Zhang-arx,Imai-PRL126}, moments of the realigned matrix \cite{Zhang-QIP21,Aggarwal-PRA109,Aggarwal-PRA108}, and their connection to principal minors \cite{Ali-QIP22,Zhang-AP534,Zhang-PRA108}. More recently, these ideas were unified and extended to moments constructed from an arbitrary positive, not necessarily completely positive, map, termed PM-moments \cite{Wang-EPJ137,Ali-QIP24}. Since the transpose map is itself a positive but not completely positive map, the PT-moments criterion arises as a special case of this more general framework. 

This work is another step in the direction of identifying and detecting bound entanglement for two-qutrits. Rather than searching for improved moment inequalities of a fixed positive map, we return to the underlying separability theorem \cite{Horodecki-PLA223}: a state $\rho_{AB}$ is entangled if and only if there exists some positive map $\Lambda$, which is not completely positive, such that $(\mathcal{I}\otimes\Lambda)(\rho_{AB})$ develops a negative eigenvalue. Since the transpose map alone only reveals NPT-entangled states, additional structural, indecomposable positive maps are needed to expose bound entanglement in $3 \otimes 3$ systems. Kye's indecomposable positive maps on $M_3(\mathcal{C})$ were among the first analytically tractable examples of such maps and, owing to their close connection to unextendible product bases, remain a benchmark for testing the strength of positive-map based separability criteria \cite{Kye-work}. Building on this line of maps, we employ an explicit structural map \cite{Hou-2011} to analyze a physically motivated three-parameter family of qutrit-qutrit states $\rho_{a,b,c}$, which contains the well-known Horodecki bound entangled family as a special case \cite{Horodecki-PRL82}.

Our main contribution is an exact, closed-form geometric characterization of the separability structure of this family, rather than a case-by-case detection statement. We derive the exact quadratic boundary of the PPT region within the physical probability simplex and identify the exact linear threshold that separates the PPT bound entangled sub-region from the remaining, most likely separable, sub-region. Applying the structural map of \cite{Hou-2011} to this decomposition, we show that a single simple criterion certifies bound entanglement over the entire region allowed by our analytical threshold, leaving no undetected gap within it.

This work is organized as follows. In section \ref{Sec:Model}, we briefly review positive maps which are not completely positive and the entanglement criterion built on them. In section \ref{Sec:BP}, we introduce the three-parameter qutrit family $\rho_{a,b,c}$ and derive the explicit geometric boundaries of its PPT and bound entangled regions, together with a numerical illustration of these regions. We summarize our results and discuss future directions in section \ref{Sec:conc}.

\section{Separability of quantum states and positive maps} 
\label{Sec:Model}

We define $\mathcal{H}_{AB} = \mathcal{H}_A \otimes \mathcal{H}_B$ as the bipartite finite dimensional Hilbert space. A mixed quantum 
state $\varrho \in \mathcal{H}_{AB}$ is a positive semidefinite density matrix with unit trace. A bipartite density matrix is said to be separable if it can be written as 
\begin{eqnarray}
\varrho^{AB} = \sum_i^n \, p_i \,  \varrho_i^A \otimes \varrho_i^B \, 
\label{Eq:sep}
\end{eqnarray}
where $p_i \geq 0 $, $\sum_i \, p_i = 1$, and $\varrho_i^A$ ($\varrho_i^B$) is a state for subsystem $A$ ($B$). If a quantum state is not separable then it is entangled. However, it is not simple to use this definition to check whether a given quantum state can be written as convex combination of product states. Despite significant efforts and some partial results, it is still an open question to decide if a given quantum state is entangled or not \cite{Horodecki-RMP81,Guehne-PR474}. 

An important contribution on this problem of separabilty is to relate the issue with theory of positive maps \cite{Horodecki-PLA223}. A map $\Lambda$ is said to be positive if it maps a positive matrix $M$ (with positive eigenvalues) to another positive matrix $N =\Lambda (M)$. A positive map is called complete positive if $\mathcal{I}_A \otimes \Lambda_B (X)$ is positive, where  $X \in \mathcal{H}_{AB}$ is positive, otherwise $\Lambda$ is called positive but not completely positive map. Such maps are quite powerful for detection of entanglement. It was found that if $\Lambda$ is a positive map but not completely positive then for a separable state $\rho_{AB}$, the matrix $\mathcal{I} \otimes \Lambda (\rho_{AB})$ must have all positive eigenvalues. This condition is both necessary and sufficient for detection of entanglement \cite{Horodecki-PLA223}. This means that if a given quantum state $\sigma_{AB}$ is entangled then there must exist a positive map $\Lambda$, such that the matrix $\mathcal{I} \otimes \Lambda (\sigma_{AB})$ will have at least one negative eigenvalue. 
Therefore, it follows that the problem of detection of entanglement is to find the positive maps which are not completely positive. This issue is non-trivial because it is not easy to find the positive maps which are not completely positive. Even if we are able to find such maps, it is not clear whether they will detect a given quantum state. Hence the challenge is to look for 'quantum state specific positive maps' as other positive maps may not detect entanglement of a given quantum state.
  
Transposition is one such positive map, which is not completely positive and hence can detect some entangled states \cite{Peres-PRL77}. A necessary condition for separability is to check the partial transpose of the density matrix. If $(\varrho^{AB})^{T_B}$ is negative (having at least one negative eigenvalue) then state $\varrho^{AB}$ is entangled. This condition is necessary and sufficient for $ 2 \otimes 2 $ and $ 2 \otimes 3$ quantum systems \cite{Horodecki-PLA223}. However, for higher dimensions of Hilbert space, there are quantum states having positive partial transpose (PPT) nevertheless entangled. In this work, we only focus on $3 \otimes 3$ quantum systems with Hilbert space having dimension $9$. It is already known that for this system, PPT-entangled states exist, however the known examples are limited. Therefore it is very important to construct and/or identify bound entangled states for this dimension of Hilbert space. Already we observe that it is not an easy task to construct positive maps to detect PPT entangled states. There are few known examples of PPT-entangled states for this system but it is not known how to construct the positive but not completely positive maps for them.  

In this work, we approach the problem from the perspective of some well known positive maps and study their range of entanglement detection. Kye’s indecomposable positive maps on $M_3(\mathcal{C})$ have played a central role in the study of bound entanglement. These maps provided some of the first systematic examples of analytically tractable, indecomposable maps capable of detecting PPT entangled states in $3 \otimes 3$ systems \cite{Kye-work}. Owing to their clear algebraic structure and close connection to unextendible product bases, Kye maps have become a benchmark for testing the strength of positive-map–based separability criteria. Let $\{|e_k\rangle\}$ with $k = 1, 2, 3$ be an orthonormal basis in $\mathcal{C}^3$. We define the elementary operators (projectors) by $E_{kl} = |e_k\rangle \langle e_l|$. (There are $9$ such operators). One particular example of a positive but not completely positive map $\Phi$ can be written as \cite{Hou-2011} 
\begin{eqnarray}
\Phi (A) = 2 \, \sum_{i=1}^{3} \, E_{ii} \, A \, E_{ii}^\dagger + E_{21} \, A \, E_{21}^\dagger 
+ E_{32} \, A \, E_{32}^\dagger \nonumber \\ +  E_{13} \, A \, E_{13}^\dagger - \bigg(\sum_{i=1}^3 E_{ii} \bigg) \, A \, 
\bigg(\sum_{i=1}^3 E_{ii} \bigg)^\dagger \, . 
\label{Eq:mp3}
\end{eqnarray}
For a positive $3 \times 3$ matrix with entries $(A)_{ij}$, this map gives us another positive matrix:
\begin{eqnarray}
\Phi (A) = \left[ 
\begin{array}{ccc}
a_{11}+a_{33} & -a_{12} & -a_{13} \\ 
-a_{21} & a_{11}+a_{22} & -a_{23} \\ 
-a_{31} & -a_{32} & a_{22}+a_{33} 
\end{array}
\right] \, .
\label{Eq:hm}
\end{eqnarray}
It was shown that the map $\Phi$ is not completely positive. We will employ this map in next section to detect bound entangled states.
 
\section{Bound entangled states for two qutrits}
\label{Sec:BP}

To systematically analyze the detection of bound entangled states, we investigate a comprehensive three-parameter family of mixed states in a qutrit-qutrit ($3 \otimes 3$) Hilbert space. Let $| \Psi\rangle = \frac{1}{\sqrt{3}}(| 11\rangle + | 22\rangle + | 33\rangle)$ represent the maximally entangled two-qutrit state. We introduce two reference noise states, $\sigma_+$ and $\sigma_-$, defined as uniform statistical mixtures of specific product states:
\begin{eqnarray}
\sigma_+ = \frac{1}{3}\big[ |12\rangle\langle 12| + | 23\rangle\langle 23| + |31\rangle\langle 31| \big] 
\end{eqnarray}
\begin{eqnarray}
\sigma_- = \frac{1}{3}\big[|21\rangle\langle 21| + | 32\rangle\langle 32| + |13\rangle\langle 13| \big]. 
\end{eqnarray}
Using these foundational components, we construct the generalized statistical ensemble:
\begin{equation}
\rho_{a,b,c} = a \, |\Psi\rangle \langle \Psi | + b \, \sigma_+ + c \, \sigma_- \,,
\end{equation}
where the parameters represent physical probabilities satisfying the conservation constraint:
\begin{equation}
a + b + c = 1, \quad \text{with } a, b, c \ge 0.
\end{equation}
This general framework contains several historically significant entanglement models. For instance, fixing $a = \frac{2}{7}$, 
$b = \frac{\alpha}{7}$, and $c = \frac{5-\alpha}{7}$ reconstructs the well-known Horodecki bound entangled family \cite{Horodecki-PRL82}.

\begin{figure}[h]
\centering
\scalebox{2.75}{\includegraphics[width=2.5in]{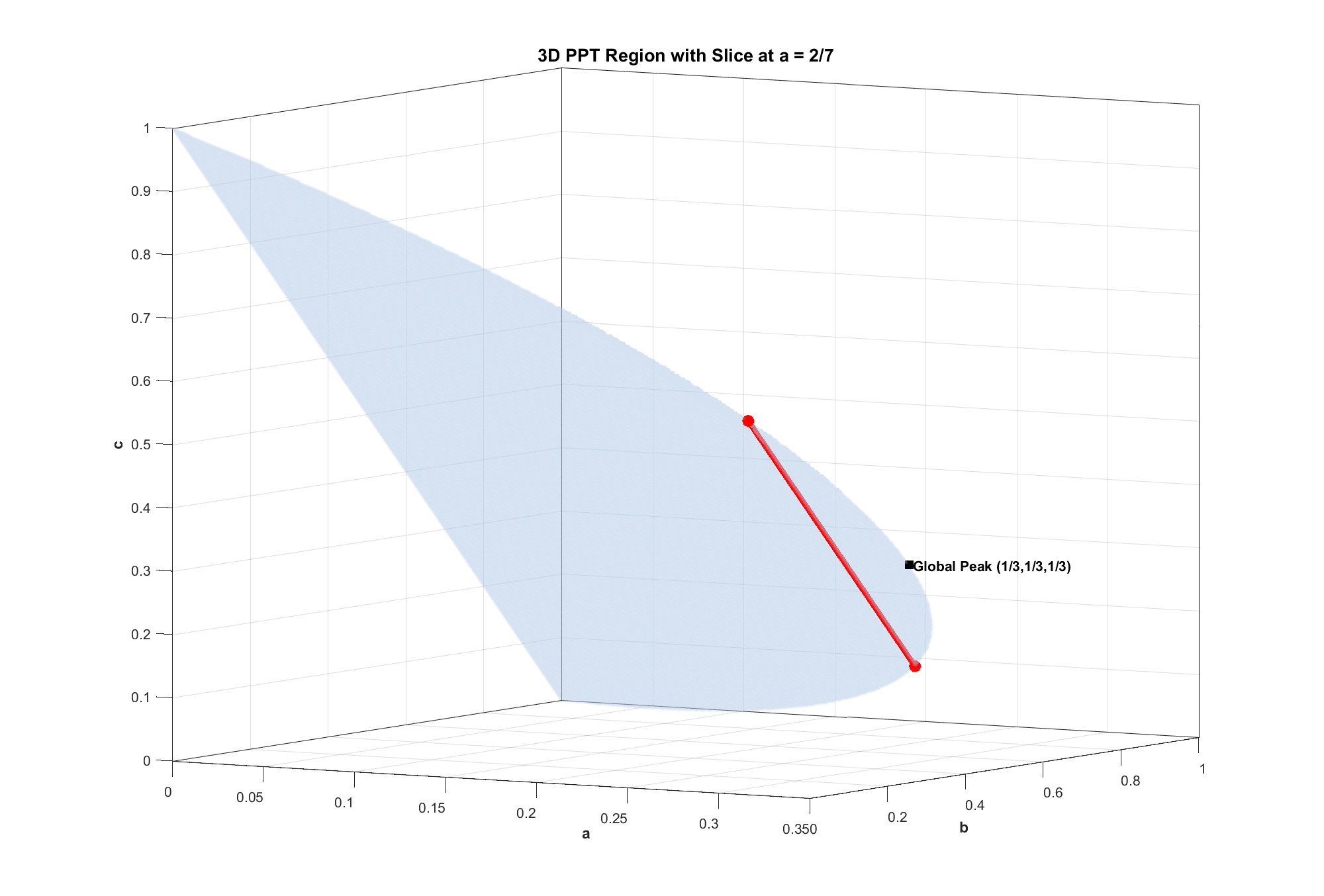}}
\caption{(Color online) The shaded leaf-like surface is the numerically obtained PPT region of $\rho_{a,b,c}$ on the probability simplex $a+b+c=1$, $a,b,c \geq 0$, satisfying $bc \geq a^2$. The global peak of the region is the symmetric point $(1/3,1/3,1/3)$. The straight (red) line marks the slice $a = 2/7$: states identified by Horodecki \cite{Horodecki-PRL82}. See text for details.}
\label{FIG:1}
\end{figure}

To isolate where bound entanglement can exist, we must first analytically map the precise geometric boundaries of the PPT region for $\rho_{a,b,c}$. By executing a partial transpose operation on the second subsystem, the state $\rho_{a,b,c}^{T_B}$ yields an explicit set of eigenvalues. Mathematical analysis shows that the state remains positive under partial transposition if and only if it satisfies the following quadratic condition:
\begin{equation}
b \, c \geq a^2 \,.
\end{equation}
To evaluate this condition inside the physical probability simplex, we eliminate the variable $c$ by substituting $c = 1 - a - b$. This converts the PPT requirement into a closed quadratic expression in terms of the remaining independent coordinates $(a, b)$:
\begin{equation}
b(1 - a - b) \geq a^2 \implies a^2 + a\,b + b^2 - b \leq 0 \, .
\end{equation}
We can establish the maximum theoretical limit of the parameter $a$ within this valid region by applying the Arithmetic Mean-Geometric Mean (AM-GM) inequality directly to the product $b\,c$:
\begin{equation}
b \, c \leq \left(\frac{b+c}{2}\right)^2 \, .
\end{equation}
Substituting the conservation identity $b+c = 1-a$ yields a strict upper bound on the product:
\begin{equation}
b\, c \leq \left(\frac{1-a}{2}\right)^2 \, .
\end{equation}
Equating this maximum geometric limit to the PPT boundary condition ($b\, c \geq a^2$) establishes the extreme constraint on the variable $a$:
\begin{equation}
a^2 \leq \left(\frac{1-a}{2}\right)^2 \implies 2a \leq 1 - a \implies 3a \leq 1 \implies a \leq \frac{1}{3} \, .
\end{equation}
For any chosen value of $a$ within this domain, the variable $b$ is strictly bounded by the roots of the corresponding quadratic characteristic equation:
\begin{equation}
\frac{1-a-\sqrt{1-2a-3a^2}}{2} \leq b \leq \frac{1-a+\sqrt{1-2a-3a^2}}{2} \,.
\end{equation}
We can numerically plot the 3D surface for the condition $b \, c \geq a^2$ with constraints $a + b + c = 1$, $a,b,c \geq 0$. In figure (\ref{FIG:1}), we have plotted the numerical solution for PPT-region (shaded region). Thus, the PPT feasible region exists as an open, leaf-like curved surface embedded directly on the oblique conservation plane $a+b+c=1$, valid exclusively for values of $a$ in the range $0 < a \leq \frac{1}{3}$. The global peak of the region is located at the symmetric point $(\frac{1}{3},\frac{1}{3},\frac{1}{3} )$. The straight (red) line represents a slice in this surface for a specific value of parameter $a = \frac{2}{7}$, which corresponds precisely to the states identified by Horodecki \cite{Horodecki-PRL82}. On this line some states are separable for $0.286 \leq b \leq 0.429 $ and PPT-entangled for $ 0.429 < b \leq 0.5714$. The NPT-states are not shown here. 

\begin{figure}[h]
\centering
\scalebox{2.5}{\includegraphics[width=2.3in]{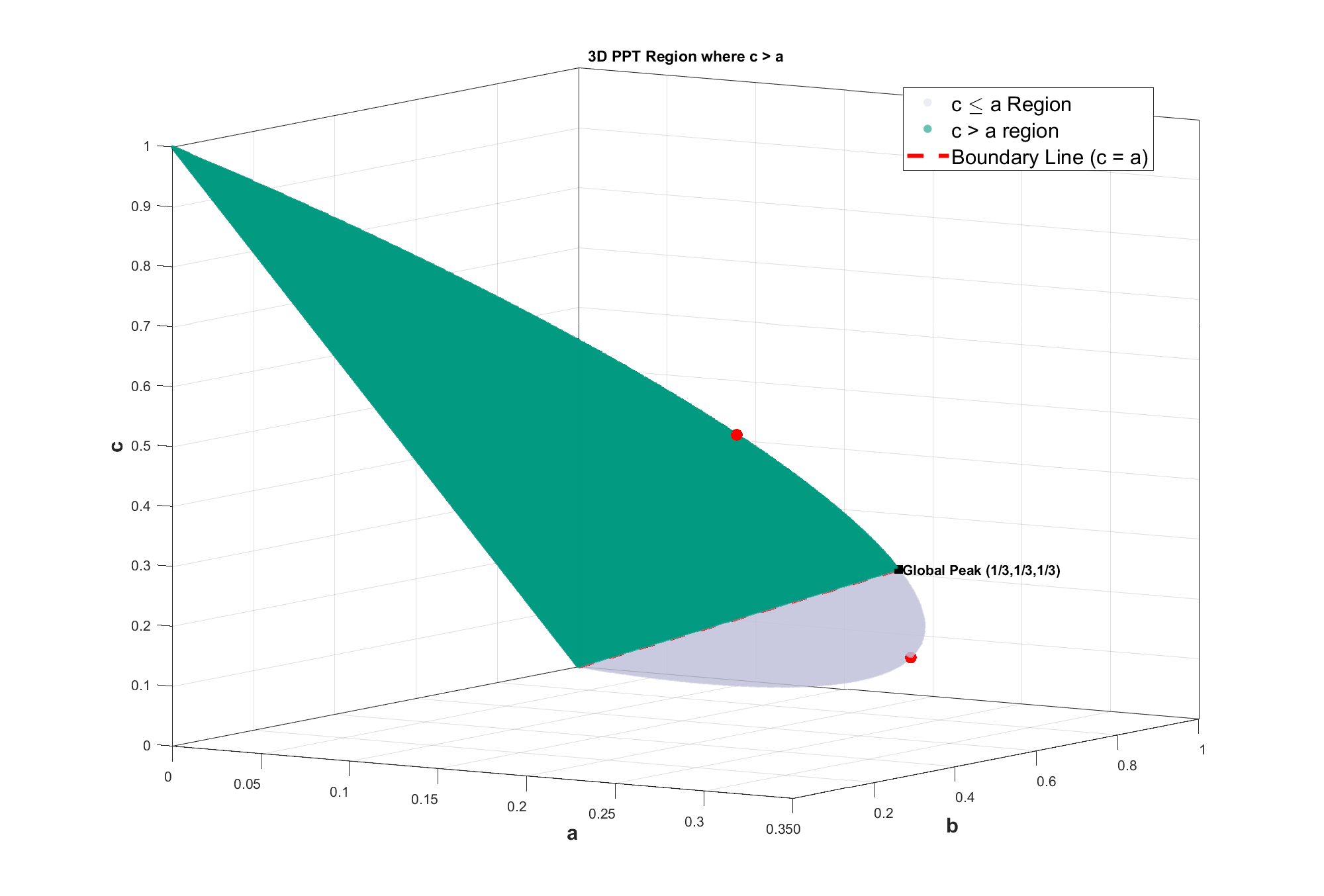}}
\caption{(Color online) Partition of the leaf-like PPT region of $\rho_{a,b,c}$ into the bound entangled sub-region $a > c$ (gray), the most likely separable sub-region $a < c$ (dark green), and the boundary line $a = c$ separating them. The structural map $\Phi$ of Eq.~(\ref{Eq:mp3}) certifies bound entanglement throughout the entire gray region. See text for more details.}
\label{FIG:2}
\end{figure}

We employ a non-completely positive linear map $\Phi: \mathcal{H}_A \to \mathcal{H}_B$ described in the previous section to the states $\rho_{a,b,c}$. This mapping is defined as follows: 
\begin{eqnarray}
\tilde{\rho}_{a,b,c} = \mathcal{I}_3 \otimes \Phi (\rho_{a,b,c}) \, . 
\label{Eq:mp3a}
\end{eqnarray}
Computing its characteristic polynomial reveals that the system generates a single isolated, potentially negative eigenvalue given by:
\begin{equation}
\lambda_{\text{critical}} = c - a \, .
\end{equation}
According to positive map theory, a state is guaranteed to be entangled if its transformed matrix develops a negative spectrum. Consequently, any state within the PPT region that satisfies the linear condition $c - a < 0$, or equivalently $a > c$, is guaranteed to be bound entangled. By rewriting this linear inequality in terms of the independent variables $(a, b)$, we get:
\begin{equation}
a > 1 - a - b \implies b > 1 - 2a \,.
\end{equation}
This introduces a sharp linear cutoff that divides the leaf-like PPT region into two distinct geometric domains. The sub-region where $b > 1 - 2a$ defines the mathematically verified PPT bound entangled domain, which terminates at the symmetric maximum coordinates $(a,b,c) = (\frac{1}{3}, \frac{1}{3}, \frac{1}{3})$. In figure (\ref{FIG:2}), we have identified three regions: the region $a > c$ (gray) consists of PPT-entangled states as indicated exclusively by positive maps; the region $a < c$ (dark green) consists of PPT states not detected by the map $\Phi$. The symmetry suggests that these are most probably separable states, because Horodecki states in this area are shown to be separable. The third region, $a = c$, is essentially the boundary line between the previous two regions, and is probably also separable. The two (red) dots represent the slice for Horodecki states. Of course, all states outside this leaf-like surface are NPT. 

We can also parametrize $a$ in terms of a denominator. Let $a = \frac{n}{d}$, where $0 < n < \frac{d}{3}$. In this case $b +c = \frac{d-n}{d}$. In addition, if we want to write $b$ and $c$ in terms of another parameter $\alpha$, then $b = \frac{\alpha}{d}$ and $c = \frac{(d-n)-\alpha}{d}$. This allows us to study various slices in this three dimensional simplex. For a specific choice $a = \frac{3}{10}$ $(n = 3, d = 10)$, the family of quantum states become
\begin{eqnarray}
\rho_\alpha = \frac{3}{10} \, |\Psi\rangle \langle \Psi | + \frac{\alpha}{10} \, \sigma_+ + \frac{7-\alpha}{10} \, \sigma_- \,,
\end{eqnarray}
where $0 \leq \alpha \leq 7$. These states have a three-fold degenerate eigenvalue that can become negative, given as
\begin{eqnarray}
\lambda_\alpha = \frac{1}{20} \, \big[ 7 - \sqrt{85-28 \, \alpha + 4 \, \alpha^2} \big] \,.
\end{eqnarray}
This eigenvalue is negative for $0 \leq \alpha < \frac{7-\sqrt{13}}{2} (\approx 1.697)$ and $\frac{7+\sqrt{13}}{2}(\approx 5.303) < \alpha \leq 7$, hence NPT entangled. For all other range $1.697 \leq \alpha \leq 5.303$, the states are PPT. However, as the parameter $c$ falls in range $0.1697 \leq c \leq 0.5303$ for PPT-region and $a = 0.3$ is fixed, therefore for all values of $ 0.1697 \leq c < 0.3$, $a$ is strictly larger than $c$, the quantum states are PPT in this range but detected by positive map $\Phi$. Hence they are bound entangled and represent another slice in the leaf-like surface. For Horodecki states $a = \frac{2}{7}$, $b = \frac{\alpha}{7}$ and $c = \frac{5-\alpha}{7}$. We have estimated that the percentage of the area detected by the positive map $\Phi$ (gray area in figure (\ref{FIG:2})) for bound entangled states is precisely $14.76 \%$, whereas the dark green area is $85.24 \%$. 

Finally, we have briefly looked into the natural extension of such states to $d \otimes d$ systems, and initial observations suggest that this case is quite different from the $3 \otimes 3$ system; we intend to study this problem in future work.  

\section{Discussion and Summary} 
\label{Sec:conc}
We have studied the entanglement properties of a family of quantum states. By utilizing a structured, non-completely positive map, we developed a verification method capable of detecting bound entanglement. Focusing on a generalized $3 \otimes 3$ qutrit mixed state family $\rho_{a,b,c}$, we derived the exact analytical boundaries for both the PPT leaf-like region ($a^2+ab+b^2-b \le 0$) and the bound entangled sub-region ($a > c$). We demonstrated that this structural map successfully exposes bound entanglement throughout the entire threshold region, hence generalizing Horodecki family of quantum states. We have shown that Horodecki states are single slice of this 3D leaf-like surface and there are much more families of such entangled states with positive partial transpose. Extending this geometric approach to other bipartite and multipartite state families is a promising direction for future work.
\subsection*{Acknowledgments}
The author is grateful to the Islamic University of Madinah for supporting this work.
%
\subsection*{Author Contributions}
I declare that this work is entirely my own, from conception through analysis and compilation.

\subsection*{Funding}
Not Applicable

\subsection*{Data Availability}
No new experimental data were generated in this study. All results are analytically derived; the figures were produced from the closed-form expressions given in the manuscript, and the underlying code can be provided by the author upon reasonable request.

\subsection*{Code Availability}
The author can provide source code used to compile the results upon request to relevant party.

\section*{Declarations}

\subsection*{Conflict of Interest}
The author declares that he has no conflict of interest.

\subsection*{Ethics approval}
Not Applicable

\subsection*{Consent to participate}
Not Applicable

\subsection*{Consent to publications}
Not Applicable


\end{document}